\documentclass[twocolumn,amsmath,amssymb,prd]{revtex4}

\def\al{\alpha}
\def\be{\beta}
\def\ga{\gamma}
\def\de{\delta}
\def\ep{\epsilon}

\def\et{\eta}
\def\th{\theta}

\def\ka{\kappa}
\def\la{\lambda}

\def\ph{\phi}

\def\ch{\chi}
\def\ps{\psi}

\def\La{\Lambda}

\def\mn{{\mu\nu}}

\def\fr#1#2{{{#1} \over {#2}}}
\def\half{{\textstyle{1\over 2}}}

\def\frac#1#2{{\textstyle{{#1}\over {#2}}}}

\def\vev#1{\langle {#1}\rangle}

\def\lsim{\mathrel{\rlap{\lower4pt\hbox{\hskip1pt$\sim$}}
    \raise1pt\hbox{$<$}}}
\def\gsim{\mathrel{\rlap{\lower4pt\hbox{\hskip1pt$\sim$}}
    \raise1pt\hbox{$>$}}}
\def\sqr#1#2{{\vcenter{\vbox{\hrule height.#2pt
         \hbox{\vrule width.#2pt height#1pt \kern#1pt
         \vrule width.#2pt}
         \hrule height.#2pt}}}}

\def\pt#1{\phantom{#1}}

\def\sss{s^{\mu\nu}}
\def\ttt{t^{\ka\la\mu\nu}}

\def\sb{\overline{s}}
\def\tb{\overline{t}}
\def\ub{\overline{u}}

\def\stw{\tilde{s}}
\def\ttw{\tilde{t}}
\def\utw{\tilde{u}}

\newcommand{\beq}{\begin{equation}}
\newcommand{\eeq}{\end{equation}}
\newcommand{\bea}{\begin{eqnarray}}
\newcommand{\eea}{\end{eqnarray}}
\newcommand{\bit}{\begin{itemize}}
\newcommand{\eit}{\end{itemize}}
\newcommand{\rf}[1]{(\ref{#1})}

\begin{document}

\title{Explicit versus Spontaneous Diffeomorphism Breaking in Gravity}

\author{Robert Bluhm}

\affiliation{
Physics Department, Colby College,
Waterville, ME 04901 
}


\begin{abstract}
Gravitational theories with fixed background fields break 
local Lorentz and diffeomorphism invariance either explicitly or spontaneously.
In the case of explicit breaking it is known that conflicts
can arise between the dynamics and geometrical constraints,
while spontaneous breaking evades this problem.
It is for this reason that in the gravity sector of the Standard-Model Extension (SME)
it is assumed that the background fields (SME coefficients) originate from spontaneous symmetry breaking.
However, in other examples, such as Chern-Simons gravity and massive gravity,
diffeomorphism invariance is explicitly broken by the background fields,
and the potential conflicts between the dynamics and geometry can be avoided.
An analysis of how this occurs is given,
and the conditions that are placed on the metric tensor and gravitational structure 
as a result of the presence of an explicit-breaking background are described.
The gravity sector of the SME is then considered for the case of explicit breaking.
However, it is found that a useful post-Newtonian limit is only obtained when the
symmetry breaking is spontaneous.
\end{abstract}


\maketitle

\section{Introduction}

The idea that local Lorentz symmetry and diffeomorphism symmetry 
might not hold exactly is frequently cited
as a key signature of new physics in quantum gravity 
and modified gravity theories that go beyond the Standard Model 
of particle physics and Einstein's General Relativity (GR)
\cite{cpt}.
The breaking of Lorentz invariance also allows breaking of the
discrete spacetime symmetry CPT,
which consists of the combination of charge conjugation,
parity, and time reversal.

Gravitational models that incorporate local Lorentz and diffeomorphism violation
at the level of effective field theory
typically do so by including fixed background fields that break the spacetime
symmetries either explicitly or spontaneously.  
Examples include vector-tensor theories motivated from string theory, 
in which local Lorentz and diffeomorphism invariance are spontaneously broken
by background vacuum expectation values,
as well as Chern-Simons gravity and massive gravity models, 
where the symmetries are explicitly broken.

Phenomenological investigations of
Lorentz and diffeomorphism violation use the theoretical framework 
known as the Standard-Model Extension (SME)
\cite{sme,akgrav04,rbsme}.
In the SME,
fixed background fields, 
referred to as SME coefficients,
couple to the matter and gravitational fields.
It is the presence of these coefficients that cause the Lorentz and diffeomorphism breaking.  
Numerous experiments have been performed in recent years
searching for violations of Lorentz symmetry, CPT, and 
diffeomorphism invariance
\cite{{LorentzTests}}.
These include gravity tests where a post-Newtonian limit of the 
SME is used as a framework in Riemann spacetime
\cite{qbak06,akjt}.
The sensitivities in these experiments are expressed as
bounds on the SME coefficients.
Extensive data tables for these bounds can be found in 
Ref.\ \cite{aknr-tables}.

In the absence of gravity,
in the context of special relativity, 
the SME coefficients can be
treated as due to either explicit or spontaneous Lorentz breaking
\cite{sme}.
However, with gravity in a curved spacetime, 
the SME coefficients are assumed to
arise from a process of spontaneous local Lorentz and diffeomorpishm breaking.
One reason supporting this view is that the original motivation for developing the SME
stemmed from the idea that Lorentz symmetry might be spontaneously broken
in the context of a quantum theory of gravity such as string theory
\cite{ks}.
At the same time,
spontaneous symmetry breaking is an elegant method that 
has wide application in physics compared to explicit symmetry breaking.
However,
it was also found by Kosteleck\'y that in the
context of gravity explicit Lorentz and diffeomorphism breaking
can lead to conflicts between the dynamics and geometrical constraints
that must hold,
while spontaneous breaking of these symmetries evades these potential conflicts
\cite{akgrav04}.
It is mainly for this reason that the gravity sector of the SME assumes the
background coefficients stem from spontaneous symmetry breaking.

In contrast, however, Chern-Simons and massive gravity models 
have background fields in their Lagrangians,
which explicitly break local Lorentz and diffeomorphism invariance.
Nonetheless, these types of models are 
for the most part able to evade the potential conflicts
between dynamics and geometrical constraints.
This therefore raises the question of what the critical differences are
in gravitational theories with fixed background fields when the symmetry
breaking is explicit versus spontaneous.
A related question is whether the gravity sector of the SME remains consistent
when the coefficients are interpreted as explicitly breaking local Lorentz and 
diffeomorphism invariance.

The main goal of this paper is to address these issues and
to examine the differences between the processes of explicit and spontaneous 
diffeomorphism breaking in gravitational theories in Riemann spacetime.  
This includes looking at the differences in interpretation and physical behavior
of the background fields that cause the symmetry breaking.
Examples of theories that are considered
include Chern-Simons gravity,
massive gravity, 
models with spontaneous Lorentz breaking,
and the gravity sector of the SME.

Traditionally, gravitational theories with fixed background fields
have been viewed as less compelling than GR.
This is because they can contain what are called 
`absolute objects' or involve `prior geometry'
\cite{A62, mtw}.
An absolute object cannot have back reactions,
although it can affect the other fields in the theory.
The term prior geometry usually implies that parts of the metric 
and background curvature are predetermined. 
In GR, features such as these do not occur,
and there are natural back reactions between the matter and gravitational fields.
This results in a direct link between geometry and the energy-momentum density. 
However, this is not necessarily the case when diffeomorphisms are broken,
although the extent of the departure from GR can depend on 
whether the symmetry breaking is explicit versus spontaneous.
Thus, a further related goal of this paper is to compare theories with
explicit or spontaneous diffeomorphism breaking with GR
and to examine whether they can retain the natural features found in GR 
or whether they are fundamentally different from GR.  

The next section begins with an overview of the different types
of symmetry breaking that are relevant.
It then gives a general overview of gravitational effective theories
with background tensors that break diffeomorphism invariance.
The potential conflicts between dynamics and geometrical constraints
in the case of explicit diffeomorphism breaking are
examined and discussed in Section III.
This is followed in Section IV by looking at specific examples of
models with fixed background fields.
Section V examines the post-Newtonian limit of the SME
and considers the possibility that the symmetry breaking is explicit
instead of spontaneous.  
A summary and conclusions are given in Section VI.

\section{Symmetry Breaking}

In theories with spacetime symmetry breaking,
it is important to make two sets of distinctions
that characterize the symmetry breaking
\cite{sme,akgrav04}.
The first is between what are
called particle and observer transformations.
The second is between 
explicit versus spontaneous symmetry breaking.

When testing a theory with fixed background fields,
it is not possible to alter the experimental setups in a way that
makes active changes in the background tensors.
The definition of particle transformations take this into account.
Under particle transformations,
dynamical tensor fields transform 
while both the background fields
and the coordinate system used to describe
the spacetime manifold are left unchanged.
On the other hand,
under observer transformations,
which are passive transformations,
all of the tensor fields
(including the background)
are left unchanged,
while the coordinate system transforms.
In the absence of symmetry breaking,
these two transformations acting on tensor components
are inversely related. 
However,
when a fixed background is present,
the particle symmetry is broken,
and the action is not invariant under the particle transformations.
Nonetheless,
a physical theory should continue to be observer invariant
even when there is a fixed background field.
Thus, the observer symmetry must continue to hold.
This requires that all of the terms in the Lagrangian 
must be scalars under observer transformations.

The particle symmetry breaking can then be characterized as either explicit or spontaneous.
If it is explicit, it is due to the appearance of the fixed
background tensor directly in the Lagrangian.
However, if it is spontaneous,
the background tensor does not initially appear in the Lagrangian
at a fundamental level,
but instead it appears in the vacuum solution for the theory.
In this case, there is a dynamical field that acquires a
vacuum expectation value,
and the full action remains invariant under both
particle and observer transformations.  

In GR,
diffeomorphisms are particle transformations,
consisting of mappings, $x^\mu \rightarrow x^\mu + \xi^\mu$,
where the changes in dynamical tensors are given by the Lie derivative,
${\cal L}_\xi$, along the direction of the vectors $\xi^\mu$.
In contrast, general coordinate transformations are observer transformations 
to a new coordinate system, $x^\mu \rightarrow x^{\mu^\prime} (x)$.
By choosing $x^{\mu^\prime}(x)$ as an infinitesimal 
coordinate transformation to $x^\mu - \xi^\mu$, 
using an opposite sign for $\xi^\mu$,
a set of observer transformations that mathematically 
have the same form as the particle diffeomorphisms can be found.
These are called observer diffeomorphisms.

In many applications in GR, it is mathematically equivalent to use 
either particle or observer diffeomorphisms,
since dynamical tensors transform the same way under either symmetry.
For example,
the components of the metric field transform as
\beq
g_{\mu\nu} \rightarrow g_{\mu\nu} + {\cal L}_{\xi} g_{\mu\nu}
= g_{\mu\nu} + D_\mu \xi_\nu + D_\nu \xi_\mu ,
\label{metricdiff}
\eeq
where $D_\mu$ is the covariant derivative.

To examine the breaking of diffeomorphism invariance,
consider a classical gravitational action with a Lagrangian density ${\cal L}$.
Symmetry breaking occurs when a fixed background tensor $\bar k_\ch$
appears in the theory,
where $\ch$ generically denotes the spacetime indices on the background tensor,
which can be contravariant, covariant, or mixed.  
If the breaking is explicit, 
$\bar k_\ch$ appears directly in the Lagrangian.
However, with spontaneous breaking
$\bar k_\ch$ does not initially appear in the Lagrangian
at a fundamental level.
Instead, it appears as a vacuum solution for the theory.
In this case, there is a dynamical field $k_\ch$ that acquires a
vacuum expectation value denoted as $\bar k_\ch = \vev{k_\ch}$,
and the full action remains diffeomorphism invariant.  

In many cases of theories with spontaneous breaking, 
particularly when perturbation theory is used,
or when massive degrees of freedom are integrated out,
it is useful to work with an effective field theory for ${\cal L}$
in which the field $k_\ch$ is truncated to its background value $\bar k_\ch$
plus small excitations about the background.
For the vacuum solution itself,
the excitations can be set to zero.
However, to maintain the diffeomorphsim invariance of the action the Nambu-Goldstone (NG)
modes for the broken symmetry would need to be included along with the vacuum value.

With these considerations in mind,
a generic action describing a gravitational effective theory
with a background field, matter fields, 
and either explicit or spontaneous diffeomorphism violation can be written 
in a low-energy limit as
\beq
S = S_{\rm EH} + S_{\rm LV} + S_{\rm LI} .
\label{act}
\eeq
In Riemann spacetime,
it is assumed that the leading gravitational contribution
to the action is the Einstein-Hilbert term of GR.  
It is given as
\beq
S_{\rm EH} = \fr {1}{2\ka} \int d^4x \sqrt{-g} \,R,
\label{EHaction}
\eeq
where $R$ is the Ricci scalar
and $\ka = 8 \pi G$.
For simplicity, a cosmological constant term has been omitted.  
The second term in $S$ contains the
Lorentz- and diffeomorphism-violating background $\bar k_\ch$,
and it has the form,
\beq
S_{\rm LV} = \int d^4 x \sqrt{-g} \, {\cal L}_{\rm LV}(g_{\mu\nu}, f^\ps, \bar k_\ch)  .
\label{LVaction}
\eeq
The Lagragian ${\cal L}_{\rm LV}$ also depends on the metric,
$g_{\mu\nu}$,
and the conventional matter fields denoted as $f^\ps$.
The spacetime label $\ps$ collectively
denotes all the component indices of the tensor $f^\ps$.
The third term contains the Lorentz- and diffeomorphism-invariant matter terms,
\beq
S_{\rm LI} = \int d^4 x \sqrt{-g} \, {\cal L}_{\rm LI} (g_{\mu\nu}, f^\ps)   .
\label{Sprimeaction}
\eeq
It includes kinetic contributions and symmetry-preserving interaction terms for $f^\ps$.

When written as an effective field theory with a fixed background tensor,
it is not immediately obvious whether the diffeomorphism breaking is
explicit or spontaneous.
However, the interpretation and behavior of the background field depends on which
type of symmetry breaking is being implemented.
For example, if the symmetry breaking is explicit,
then there are no excitations in the background fields $\bar k_\ch$.
The background is fixed  and there are no dynamical field variations of it in the action. 
On the other hand, if the breaking is spontaneous,
then $\bar k_\ch$ is the vacuum value of a dynamical field $k_\ch$.
The full theory contains additional excitations of $k_\ch$ around $\bar k_\ch$,
including the NG modes.
However, the effective theory in terms of $\bar k_\ch$ alone
is a truncated version of the full theory.
It has vacuum solutions as exact solutions, 
or it can be viewed as a gauge-fixed theory where the 
diffeomorphism symmetry is hidden as opposed to being explicitly broken. 

Regardless of the type of particle symmetry breaking,
each term in the action $S$ is invariant under general coordinate transformations,
including the special case of observer diffeomorphisms.
For these transformations,
all of the fields $g_{\mu\nu}$, $f^\ps$, and $\bar k_\ch$ transform using
the Lie derivative.
For example,
\beq
\bar k_\ch \stackrel {\rm obsv} { \longrightarrow} \bar k_\ch + {\cal L}_{\xi} \bar k_\ch ,
\label{kbarobsvtrans}
\eeq
under observer diffeomorphisms,
where the expression for the Lie derivative depends on the type
and number of indices that $\bar k_\ch$ has.

In contrast, the term $S_{\rm LV}$ is not invariant under particle diffeomorphisms.
These are broken by the background field $\bar k_\ch$,
which mathematically obeys ${\cal L}_{\xi} \bar k_\ch \ne 0$ for
the broken diffeomorphisms $\xi^\mu$.
Here, the breaking occurs because the field $\bar k_\ch$ remains a fixed 
background under particle diffeomorphisms and does not transform
along with the other fields in the theory.
Its transformation rule is therefore
\beq
\bar k_\ch \stackrel {\rm part} { \longrightarrow} \bar k_\ch 
\label{kbarparticletrans}
\eeq
under particle diffeomorphisms.

In the effective field theory, 
when the symmetry breaking is spontaneous,
and the field $k_\ch$ acquires a vacuum value,
$\bar k_\ch = \vev{k_\ch}$,
then an exact solution must involve
a corresponding vacuum solution
for the metric, $\bar g_{\mu\nu} = \vev{g_{\mu\nu}}$.
The ordinary background fields are assumed for simplicity 
to have vanishing vacuum values, $\vev{f^\ps} = 0$.
Although $\bar k_\ch$ is a fixed background, 
when the symmetry breaking is spontaneous it is still a solution of
the $k_\ch$ equations of motion and therefore obeys
\beq
 \int d^4x 
\sqrt{-g} \fr {\de {\cal L}_{\rm LV}} {\de \bar k_\ch} \de \bar k_\ch
= 0 
\quad ({\rm spontaneous})
\label{ksponeq}
\eeq
when the other fields take their vacuum values,
$g_{\mu\nu} = \bar g_{\mu\nu}$
and $\vev{f^\ps} = 0$.

On the other hand,
when the symmetry breaking is explicit, 
then the fixed background  
$\bar k_\ch$ represent a nondynamical field,
which does not have equations of motion and can therefore obey
\beq
 \int d^4x 
\sqrt{-g} \fr {\de {\cal L}_{\rm LV}} {\de \bar k_\ch} \de \bar k_\ch
\ne 0 
\quad ({\rm explicit}) .
\label{kexeq}
\eeq

\section{Explicit Breaking}

General features of gravitational theories with local Lorentz and diffeomorphism breaking
were investigated in Ref.\ \cite{akgrav04}.
For the case of explicit symmetry breaking, 
potential inconsistencies were found,
which were shown not to occur when the symmetry breaking is spontaneous. 
The formalism in \cite{akgrav04} uses a vierbein description 
in order to include fermions in a gravitational theory.  
A vierbein treatment is also useful in that it reveals that 
when local Lorentz symmetry is broken
by a background field so is diffeomorphism invariance
\cite{rbak}.
It allows for dynamical torsion in addition to curvature in 
a geometry that is Riemann-Cartan
\cite{fins}.
The potential incompatibility for the case of explicit breaking found in \cite{akgrav04}
involves considering the dynamical equations of motion,
the conservation laws for energy-momentum and spin density, 
and geometric identities for the curvature and torsion.

In this section, the potential conflicts stemming from
explicit diffeomorphism breaking are examined in further detail
for gravitational theories restricted to Riemann spacetime.
In this case, and in the absence of fermions,
there is no torsion and a vierbein is not needed.
In this restricted treatment, the action takes the generic form as given in Eq.\ \rf{act}.

\subsection{Consistency Requirements}

With explicit symmetry breaking,
the tensor $\bar k_\ch$ is a fixed background,
which is nondynamical and does not transform under
particle diffeomorphisms.
The action term $S_{\rm LI}$ therefore only involves conventional matter terms
and the metric,
and its variation with respect to the metric gives a
contribution to the energy-momentum tensor,
which can be labeled as
$T^{\mu\nu}_{\rm LI}$.

The symmetry-breaking term $S_{\rm LV}$ in general can
involve derivative functions of the metric and the connection,
such as the curvature tensor or its contractions.
Variation of this term with respect to the metric involves using
integration by parts.
Assuming contributions from the boundaries vanish,
these variations can be written as
\bea
\de S_{\rm LV} &=& 
 \int d^4x 
 \fr {\de (\sqrt{-g} {\cal L}_{\rm LV})} {\de g_{\mu\nu}} \de g_{\mu\nu} 
 \nonumber
 \\
& \equiv&  \fr 1 2  \int d^4x 
 \sqrt{-g} \, T^{\mu\nu}_{\rm LV} \de g_{\mu\nu}  ,
 \label{geq1}
 \eea
which defines a quantity denoted as $T^{\mu\nu}_{\rm LV}$.

Variation of the full action gives Einstein's equations,
 \beq
G^{\mu\nu} = \ka \left( T^{\mu\nu}_{\rm LI} + T^{\mu\nu}_{\rm LV} \right) ,
\label{einstein}
\eeq
where $G^{\mu\nu}$ is the Einstein tensor.
The remaining equations of motion are obtained by
variation of the conventional matter fields $f^\ps$,
which gives
 \beq
  \int d^4x 
\sqrt{-g}
\left( \fr {\de  {\cal L_{\rm LI}}} {\de f^{\ps}} +  \fr {\de  {\cal L_{\rm LV}}} {\de f^{\ps}}
\right) \de f^\ps  = 0 .
\label{feqs1}
\eeq

Taking the covariant divergence of Einstein's equations
and using the contracted Bianchi identity,
$D_\mu G^{\mu\nu} = 0$,
gives an on-shell condition that must hold,
\beq
D_\mu (T^{\mu\nu}_{\rm LI} + T^{\mu\nu}_{\rm LV}) = 0 .
\label
{Tcons}
\eeq

Next, consider the variation of $S$ with respect to observer diffeomorphisms.
Although these variations are not physically significant on their own,
mathematically they can be combined with dynamical field variations to
give results that are meaningful.  
Under observer diffeoomorphisms all three sets of fields
(including $\bar k_\ch$) transform mathematically,
where the variations are given by the Lie derivatives,
$\de g_{\mu\nu} = {\cal L}_\xi g_{\mu\nu}$, 
$\de f^\ps = {\cal L}_\xi f^\ps$, 
and $\de \bar k_\ch = {\cal L}_\xi \bar k_\ch$.
Since $S$ is invariant under general coordinate transformations,
including observer diffeomorphisms,
it follows that $(\de S)_{\rm observer} = 0$
under these transformations.
At the same time, with appropriate boundary conditions,
these observer diffeomorphism variations for $g_{\mu\nu}$ and $f^\ps$ 
are subsets of the dynamical field variations, $\de g_{\mu\nu}$ and $\de f^\ps$.
This allows the dynamical equations of motion to be combined with
the observer diffeomorphism transformations.

In addition to overall invariance,
each individual term in \rf{act}
is an observer scalar as well.
When the variations $\de g_{\mu\nu} = {\cal L}_\xi g_{\mu\nu}$
are applied to the Einstein-Hilbert term,
and integration by parts is used,
the result is the contracted Bianchi identity,
which vanishes.
Performing observer diffeomorphism variations on 
the term with ${\cal L}_{\rm LI}$ gives the condition
\beq
 \int d^4x 
 \left( 
 \fr {\de (\sqrt{-g} {\cal L}_{\rm LI})} {\de g_{\mu\nu}} {\cal L}_\xi g_{\mu\nu}
 + \sqrt{-g}  \, \fr {\de  {\cal L}_{\rm LI}} {\de f^{\ps}} {\cal L}_\xi f^\ps
 \right) = 0.
 \label{obsLI}
 \eeq
Making similar observer variations in the term with ${\cal L}_{\rm LV}$ gives
\bea
 \int d^4x 
 \left[ 
 \fr {\de (\sqrt{-g} {\cal L}_{\rm LV)}} {\de g_{\mu\nu}} {\cal L}_\xi g_{\mu\nu}
 \right.
 \quad\quad \quad\quad \quad\quad \quad\quad \quad\quad
 \cr
 \left.
  + \sqrt{-g}  \, 
  \left(
  \fr {\de  {\cal L}_{\rm LV}} {\de f^{\ps}} {\cal L}_\xi f^\ps
 +  \fr {\de  {\cal L}_{\rm LV}} {\de \bar k_\ch} {\cal L}_\xi \bar k_\ch
 \right) \right] = 0.
 \label{obsLV}
 \eea
Adding \rf{obsLI} and \rf{obsLV},
integrating by parts for the terms involving 
${\cal L}_\xi g_{\mu\nu} = D_\mu \xi_\nu + D_\nu \xi_\mu$,
and using the dynamical equations \rf{feqs1} for the matter fields
gives the relation,
\beq
 \int d^4x \sqrt{-g}  \, 
 \left[ D_\mu \left( T^{\mu\nu}_{\rm LI} + T^{\mu\nu}_{\rm LV} \right) \xi_\nu
 -  \fr {\de  {\cal L}_{\rm LV}} {\de \bar k_\ch} {\cal L}_\xi \bar k_\ch
 \right] = 0.
 \label{combine1}
 \eeq
 This equation must hold due to the combination of
 general covariance, the Bianchi identity, and the
 dynamical equations of motion.

In contrast, under particle diffeomorphism transformations,
when there is explicit breaking
the symmetry does not hold,
and $(\de S)_{\rm particle} \ne 0$.
However, the only difference mathematically
between the particle diffeomorphism
transformations and the observer transformations is that the
variations involving ${\cal L}_\xi \bar k_\ch$ are missing 
in $(\de S)_{\rm particle}$ because 
$\bar k_\ch$ does not transform under the broken particle diffeomorphisms.
Combining this result with \rf{combine1} allows the condition 
for diffeomorphism violation to be written as:
\beq
(\de S)_{\rm particle} =  
\int d^4x \sqrt{-g}  \,
 \fr {\de  {\cal L}} {\de \bar k_\ch} {\cal L}_\xi \bar k_\ch \ne 0 .
 \label{particleLV1}
 \eeq
 
The potential conflict between the dynamics and the contracted Bianchi identity 
for the case of explicit diffeomorphism breaking is then readily apparent.
In particular, if \rf{Tcons} is applied on-shell in \rf{combine1},
the result is
\beq
 \int d^4x \sqrt{-g}  \,
 \fr {\de  {\cal L}} {\de \bar k_\ch} {\cal L}_\xi \bar k_\ch = 0 .
 \label{conflictcondition}
 \eeq
This appears to be in conflict with the condition of explicit breaking in Eq.\ \rf{particleLV1}.
That is, unless the integral in both \rf{particleLV1} and \rf{conflictcondition} vanishes on shell,
the resulting gravity theory with explicit diffeomorphism breaking is inconsistent.

Notice that for the case of spontaneous diffeomorphism breaking,
the interpretation of the background fields $\bar k_\ch$ as
vacuum expectation values of a dynamical field gives the condition
in \rf{ksponeq}.
Therefore,
the integral in \rf{particleLV1} and \rf{conflictcondition} vanishes on shell
and the potential conflict is avoided for spontaneous breaking.  

\subsection{Extra Degrees of Freedom}

The source of the potential inconsistency in Riemann spacetime stems from the fact that
a theory with explicit diffeomorphism breaking must still be covariant under
general coordinate transformations. 
Since observer diffeomorphisms are special cases of general coordinate
transformations and have the same mathematical form as the
broken particle diffeomorphisms,
this would seem problematic.
However,
the number of independent degrees of freedom changes as well when
diffeomorphisms are explicitly broken,
and the structure of the dynamical equations of motion that apply 
when the theory is on shell is altered.

In GR,
in addition to the matter-field degrees of freedom,
the metric $g_{\mu\nu}$ has only two physical propagating degrees of freedom.
This follows because in the Einstein-Hilbert action,
four of the ten metric components can be
shown to be auxiliary fields that do not propagate as physical modes.
This leaves at most six independent propagating modes for the metric.
In addition, with diffeomorphism invariance,
four of the metric components are gauge degrees of freedom,
which can be eliminated using the four local symmetry transformations
given in terms of the vectors $\xi^\mu$.
Thus, only two of the metric modes remain as physical degrees of freedom.  

However, in theories where diffeomorphism invariance is explicitly broken
by the presence of a fixed background,
the action is no longer invariant under these transformations.
As a result, there are four extra components in the metric
compared to theories where diffeomorphism invariance holds.
These four extra metric components are the degrees of freedom 
that would be gauge if diffeomorphism invariance were not broken.

Further insight can be gained by examining the structure of the gravitational 
equations of motion.
First, consider the ten Einstein equations,
$G^{\mu\nu} = \ka T^{\mu\nu}$.
One way of interpreting these equations is that
with the contracted Bianchi identity,
$D_\mu G^{\mu\nu} = 0$, there are effectively only six components
in $G^{\mu\nu}$ that are dynamically independent and 
act as kinetic terms for the metric $g_{\mu\nu}$.
This coincides with the maximal number of physical propagating
degrees of freedom in $g_{\mu\nu}$.
The remaining four equations, obtained after taking the covariant derivative,
reduce to the condition of covariant energy-momentum conservation,
$D_\mu T^{\mu\nu} = 0$.
However, in a theory with diffeomorphism invariance,
these equations are not satisfied using any of the six potentially physical
degrees of freedom in $g_{\mu\nu}$.
Instead, they are directly linked to the matter field equations of motion and hold
on shell as a result of the matter dynamics.
For example, 
in the absence of matter,
$T^{\mu\nu} = 0$,
and the equations are trivially satisfied.

On the other hand,
in a theory where diffeomorphism symmetry is explicitly violated,
there are again ten Einstein equations as shown in \rf{einstein}.
Because of the contracted Bianchi identity, 
it is still possible to make a distinction 
between six of these equations being associated
with the six dynamically independent components in $G^{\mu\nu}$,
while the remaining four reduce to the requirement given in \rf{Tcons}.
However, unlike a theory with diffeomorphism invariance,
the background $\bar k_\ch$ is not dynamical
and it cannot cause the equations in \rf{Tcons} to hold.
Furthermore, in the absence of conventional matter, 
$T^{\mu\nu}_{\rm LI} = 0$,
while the tensor $T^{\mu\nu}_{\rm LV}$ in \rf{Tcons}
does not vanish.
Thus,
the condition $D_\mu T^{\mu\nu}_{\rm LV} = 0$ must hold
in the absence of conventional matter.  
The only remaining degrees of freedom that can be used
to account for these equations are the four extra would-be-gauge 
degrees of freedom that appear in the 
metric when diffeomorphisms are explicitly broken.  

\subsection{Role of Extra Degrees of Freedom}

It is straightforward to show that the leading-order effect of the four extra metric 
modes that appear with explicit breaking is that they
impose \rf{Tcons} as their equations of motion.
To see this,
introduce a field redefinition for the metric, $g_{\mu\nu}$,
dividing it into six degrees of freedom denoted as $\tilde g_{\mu\nu}$
and four would-be-gauge degrees of freedom defined
in terms of a vector $\Xi_\mu$.
This gives
\beq
g_{\mu\nu} = \tilde g_{\mu\nu} + D_\mu \Xi_\nu + D_\nu \Xi_\mu .
\label{gZe}
\eeq
In this redefinition,
$\tilde g_{\mu\nu}$ consists of ten fields obeying four conditions.
Effectively, it is a gauge-fixed form of the metric,
which is not equivalent to $g_{\mu\nu}$ because of the
diffeomorphism breaking.
At leading order in $\Xi_\mu$, the covariant derivatives in
this expression can be computed using $\tilde g_{\mu\nu}$.
Using this field redefinition in the action \rf{act}
and performing the field variations $\de \Xi_\mu$ with respect to $\Xi_\mu$ 
gives the equations of motion,
\beq
\de S = \int d^4 x \, \fr {(\sqrt{-g} {\cal L})}{\de g_{\mu\nu}} 
(D_\mu \de \Xi_\nu + D_\nu \de \Xi_\mu) = 0 .
\label{Xieq}
\eeq
Substituting for ${\cal L}$
and integrating by parts gives the result,
\beq
\int d^4 x \sqrt{-g} \, \left( D_\mu G^{\mu\nu}
- \ka D_\mu (T^{\mu\nu}_{\rm LI} + T^{\mu\nu}_{\rm LV}) \right) \de \Xi_\mu = 0 .
\label{Xieq2}
\eeq
Combining this with the contracted Bianchi identity,
shows that at leading order \rf{Tcons} holds as a result of the equations of motion for the 
would-be-gauge components in the metric defined in terms of $\Xi_\mu$.

Note that even in a diffeomorphism-invariant theory
a similar argument can be made involving the gauge degrees of freedom.
In this case, the result that would follow is that variation of the action with respect to the gauge components in the 
metric leads to the requirement of covariant energy-momentum conservation.
However,
in diffeomorphism-invariant theories,
when the field redefinition involving \rf{gZe} is made,
there are always compensating field redefinitions that can be made in
the matter fields as well,
such that the gauge degrees of freedom drop out.
In this case, the condition of covariant energy-momentum conservation stems
from the dynamics of the matter fields that are not gauge related,
and there are no physically distinct solutions that arise from variation
with respect to the gauge modes.

Also note that it is not sufficient if $\Xi_\mu$ only gives rise to the condition in \rf{Tcons}
as an equation of motion.
The condition itself needs to be satisfied as well
without requiring that the background $\bar k_\ch$ must vanish.
If it turns out that the full expression for the divergence of $T^{\mu\nu}_{\rm LV}$
does not itself involve the would-be-gauge modes in the metric,
then nontrivial solutions might not exist.
For example, when using an ansatz form for the metric,
such as in cosmology, this might be a problem.
If an assumed ansatz suppresses too many of the would-be-gauge modes
it may not be possible for the condition in \rf{Tcons} to hold for 
nonzero values of the background $\bar k_\ch$.
Similarly if $T^{\mu\nu}_{\rm LV}$ is constructed out of tensors
that are invariant under infinitesimal diffeomorphisms,
such as the curvature tensor in linearized gravity,
then the fields $\Xi_\mu$ might not appear except at higher order.
In this case, perturbative treatments can become problematic.  

In a linearized treatment in a flat background,
the extra metric modes also drop out of the 
linearized Einstein tensor, $G^{\mu\nu}$.
This suggests that there are no kinetic terms for these modes at leading order
unless such terms are generated by the interactions with the fixed background.
However, even if kinetic terms do arise for the extra metric modes,
they would likely appear with an indefinite sign and could therefore behave as ghost modes.
Assuming as well that the ordinary matter sector is covariantly coupled with the metric,
there will not be any interactions at leading order between conventional matter particles
and the four additional metric modes.
Thus, the most likely scenario appears to be that the 
extra metric modes are limited to behaving
as auxiliary degrees of freedom that impose covariant energy-momentum conservation
as their equations of motion,
but which otherwise lack a physical existence of their own.  

It is important to keep in mind, however, 
that the interaction terms involving the fixed background can also
generate mass terms for the metric modes that do not have the form of gauge excitations.
In this case, kinetic terms for these modes can arise in the Einstein tensor.
As a result, 
modifications to the propagation of gravitational waves 
in theories with explicit diffeomorphism breaking should be expected.
How this occurs and how many physical massive metric modes
can propagate depends on the detailed form of the theory.
The question of whether specific choices of models are ghost free 
is then of primary importance in this context.  

\subsection{Potential Terms}

In many cases of interest,
the symmetry-breaking term ${\cal L}_{\rm LV}$ takes the form of
an interaction term involving only the metric (without derivatives)
and the background $\bar k_\ch$.
The matter fields $f^\ps$ do not couple to $\bar k_\ch$.
In this case,
${\cal L}_{\rm LV}$ is a potential term and can be written as
\beq
{\cal L}_{\rm LV} = - {\cal U} (g_{\mu\nu}, \bar k_\ch) .
\label{calU}
\eeq
An example of this form is massive gravity where the mass terms involve 
only the contractions of the metric and a fixed background field.

With the conventional matter fields only appearing in ${\cal L}_{\rm LI}$,
the tensor $T^{\mu\nu}_{\rm LI}$ becomes
the matter energy-momentum tensor.
It is assumed to be conserved
and therefore obeys $D_\mu T^{\mu\nu}_{\rm LI} = 0$ on shell.
The symmetry-breaking contribution,  $T^{\mu\nu}_{\rm LV}$, in this case can
be interpreted as the energy-momentum tensor for the background field $\bar k_\ch$.
It is conserved only if the consistency requirements can be nontrivially resolved.

With these restrictions, the consistency analysis simplifies.
Performing variations consisting of infinitesimal observer diffeomorphisms
on ${\cal U}$,
which is a scalar, 
the condition in \rf{combine1} can be shown to reduce to 
\beq
 \int d^4x \sqrt{-g}  \, 
 \left[ (D_\mu T^{\mu\nu}_{\rm LV}) \xi_\nu
 +  \fr {\de  {\cal U}} {\de \bar k_\ch} {\cal L}_\xi \bar k_\ch
 \right] = 0.
 \label{Ueq2}
 \eeq
 Here, the variations with respect to $\bar k_\ch$ need not vanish
 because $\bar k_\ch$ is not dynamical,
 and ${\cal L}_\xi \bar k_\ch$ need not vanish either because $\bar k_\ch$
 explicitly breaks diffeomorphisms.  
 These two conditions would appear to prevent $D_\mu T^{\mu\nu}_{\rm LV} = 0$
 from holding on shell.
 
 However, it can be shown in general that the two terms 
 in the integrand in \rf{Ueq2} differ by a total derivative.
 In that case, when $D_\mu T^{\mu\nu}_{\rm LV} = 0$ holds on shell,
 the remaining integral becomes a surface term.
 The condition
 \beq
 \int d^4x \sqrt{-g}  \, 
 \fr {\de  {\cal U}} {\de \bar k_\ch} {\cal L}_\xi \bar k_\ch = 0 
 \label{Ueq3}
 \eeq
then holds on shell despite the fact that the integrand does not vanish.
To show this,
first consider the definition of $T^{\mu\nu}_{\rm LV}$.
Since there are no derivatives involving $g_{\mu\nu}$ in ${\cal U}$,
standard Euler-Lagrange variations can be used, giving
\beq
T^{\mu\nu}_{\rm LV} = - g^{\mu\nu} {\cal U} - 2 \fr {\de {\cal U}} {\de g_{\mu\nu}} .
\label{TU}
\eeq
Taking a divergence of this and multiplying by $\xi_\nu$ gives
\beq
(D_\mu T^{\mu\nu}_{\rm LV}) \xi_\nu = - \xi^\mu D_\mu {\cal U} 
+ D_\mu (2 \fr {\de {\cal U}} {\de g^{\al\be}} g^{\mu\al} g^{\be\nu}) \xi_\nu .
\label{DTU}
\eeq
Combining this with the expression for the Lie derivative along $\xi^\mu$
acting on ${\cal U} (g_{\mu\nu}, \bar k_\ch)$ gives
the off-shell result 
\beq
(D_\mu T^{\mu\nu}_{\rm LV}) \xi_\nu 
+ \fr {\de  {\cal U}} {\de \bar k_\ch} {\cal L}_\xi \bar k_\ch
= D_\mu \left(2 \fr {\de {\cal U}} {\de g^{\al\be}} g^{\mu\al} \xi^\be \right)  .
\label{DTU2}
\eeq
Thus, on shell when $D_\mu T^{\mu\nu}_{\rm LV} = 0$,
the condition in \rf{Ueq3} does indeed hold,
since the integrand becomes a total derivative.

This is a general result that holds for a large class of theories with a potential ${\cal U}$.
However, there can be exceptions.
If, for example, the total divergence in \rf{DTU2} vanishes
for certain values of $g_{\mu\nu}$, $\bar k_\ch$ and $\xi^\mu$,
while the second term on the left-hand side does not,
then an inconsistency can still arise.
In this case, the only resolution would be that the background $\bar k_\ch$ must vanish.

Alternatively,
if ${\cal L}_\xi \bar k_\ch = 0$ holds for a subset of transformations
with vectors $\xi^\mu$,
then the condition in \rf{DTU2} shows that on shell
both terms on the left-hand side would vanish.
Therefore, the total covariant divergence on the right-hand side would vanish on shell as well.
Multiplying by $\sqrt{-g}$ gives the result,
\beq
\partial_\mu \left(2 \sqrt{-g}  \fr {\de {\cal U}} {\de g^{\al\be}} g^{\mu\al} \xi^\be \right) = 0 .
\label{currents}
\eeq
In a gravitational theory,
these equations impose conditions on the metric $g_{\mu\nu}$,
which can lead to restrictions on the allowed geometry of the theory.
In particular,
in approaches using an ansatz form for the metric and background tensor,
it may turn out that the condition in \rf{currents}  is restrictive enough to
rule out specific types of solutions.
 
\section{Examples}

This section looks at specific examples of gravitational theories
with fixed background fields in Riemann spacetime.
In each case,
an examination is made concerning how the potential
conflicts between dynamics and geometrical identities
are either evaded or not evaded.  

\subsection{Spacetime-Dependent Cosmological Constant}

A simple illustration of when a fatal conflict arises is
provided by the case of a gravitational theory with a
prescribed spacetime-dependent cosmological constant $\La (x)$.
Adding such a term to the Einstein-Hilbert action 
with conventional matter gives
\beq
S = \int d^4 x \sqrt{-g} \, 
\left[ \fr 1 {2\ka} (R - 2 \La(x) ) + {\cal L}_{\rm M} \right].
 \label{cosmot}
 \eeq
Here, 
the fixed background is a scalar $\bar k_\ch = \La(x)$,
${\cal L}_{\rm LI} = {\cal L}_{\rm M} $
for the matter sector,
and the symmetry-breaking term is identified as
${\cal L}_{\rm LV} = - \La(x)/\ka$.

For this theory, with $\La (x) \ne 0$,
particle diffeomorphisms are explicitly broken
and the change in the action under these transformations in \rf{particleLV1} 
can be verified to be nonzero.
First,
the variation of ${\cal L}_{\rm LV}$ with respect to 
$\bar k_\ch$ equals $-1/\ka$ and is nonzero.
Second,
the Lie derivative acting on $\La(x)$ gives
${\cal L}_\xi \La(x) =  \xi^\mu \partial_\mu \La (x)$,
which also does not vanish.
Combining these confirms \rf{particleLV1}.

However, Einstein's equations in this case are
$G^{\mu\nu} = - \La (x) g^{\mu\nu} + \ka T^{\mu\nu}_{\rm M} $,
where $T^{\mu\nu}_{\rm M} $ 
is the energy-momentum for matter and
$T^{\mu\nu}_{\rm LV} = - g^{\mu\nu} \La (x)/\ka $.
Taking the divergence of Einstein's equations,
using the contracted Bianchi identity
and $D_\mu T^{\mu\nu}_{\rm M} =0$, 
gives the condition
$D_\mu T^{\mu\nu}_{\rm LV} = - (1/\ka) g^{\mu\nu} \partial_\mu \La (x) = 0$.
Acting on the latter equation with $g_{\nu\al}$ shows that
$\partial_\al \La (x) = 0$ must hold.
This clearly contradicts the assumption that the theory
has explicit breaking and $\La (x) \ne 0$.
The theory is therefore inconsistent,
with the only acceptable resolution being that $\La$ must be constant,
which restores the diffeomorphism invariance
and \rf{particleLV1} no longer applies.

Note that in this example there are no solutions
involving would-be-gauge modes for the metric.
The scalar background $\La (x)$ does not couple to
the metric in a way that allows 
these extra degrees of freedom
to step in and impose the condition of energy-momentum conservation.
Instead, these equations involve only the background $\La (x)$,
which has no dynamics.

\subsection{Chern-Simons Gravity}

The Chern-Simons term was originally introduced in three-dimensional
gauge field theory and gravity models
\cite{djt}.
The possibility of modifying a four-dimensional gravity theory was
subsequently considered as well
\cite{rjsp}.
Its construction involves introducing a prescribed nondynamical
scalar $\th(x)$ and an associated embedding coordinate, 
$v_\mu = \partial_\mu \th \ne 0$,
which explicitly break diffeomorphism invariance
\cite{CSnote}.  

One form of the action for four-dimensional Chern-Simons gravity
can be written as \cite{rjsp} 
\beq
S_{\rm CS} 
=  \int d^4x \left(  \fr 1 {2\ka} ( \sqrt{-g} R + \fr 1 4 \th {^*R}R  )
+ \sqrt{-g} {\cal L}_{M} \right) .
 \label{CS}
 \eeq
 Here,
${^*R}R = \, ^*R^{\ka \pt{\la} \mu\nu}_{\pt{\ka}\la} R^{\la}_{\pt{\la}\ka\mu\nu}$
is the gravitational Pontryagin density,
with $^*R^{\ka \pt{\la} \mu\nu}_{\pt{\ka}\la} = \half \ep^{\mu\nu\al\be} R^{\ka}_{\pt{\ka}\la\al\be}$.
In this context, the fixed background $\bar k_\ch$ becomes the prescribed scalar $\th(x)$.
The energy-momentum $T_{\rm M}^{\mu\nu}$ stemming from the matter term ${\cal L}_{M}$ 
is assumed to be conserved,
obeying $D_\mu T_{\rm M}^{\mu\nu} = 0$.

The variation of the Chern-Simons action with respect to the metric
gives Einstein's equations,
\beq
G^{\mu\nu} + C^{\mu\nu} = \ka T_{\rm M}^{\mu\nu} ,
\label{CSEinstein}
\eeq
where $C^{\mu\nu}$ is the four-dimensional analogue of the Cotton tensor.
Here,  $C^{\mu\nu} = \ka T_{\rm LV}^{\mu\nu}$ takes the place of the 
tensor associated with the symmetry-breaking term.
The divergence of $C^{\mu\nu}$ can be computed explicitly,
giving the result \cite{rjsp},
\beq
D_\mu C^{\mu\nu} = \fr 1 {8 \sqrt{-g}} (D^\nu \th) \, {^*R}R .
\label{DCmunu}
\eeq

To examine the symmetry breaking,
variation of the Chern-Simons action under particle diffeomorphisms
can be performed as in \rf{particleLV1}.
The result is 
\beq
(\de S)_{\rm particle} 
=   \int d^4x \sqrt{-g}  \,
 \fr 1 4 {^*R}R \, \xi^\mu D_\mu \th .
 \label{CSparticleLV}
 \eeq
 Explicit diffeomorphism breaking occurs when $D_\mu \th \ne 0$
 and the Pontryagin density does not vanish.
 
 The conflict with dynamics and geometry 
 occurs when Einstein's equations  
 are combined with the contracted Bianchi identity.
 Taking a covariant derivative in \rf{CSEinstein} gives that 
 $D_\mu C^{\mu\nu} = 0$ must hold on shell.
 This requires that the product ${^*R}R \,  (D^\nu \th)$ must vanish on shell.
 Thus, either $D^\nu \th = 0$, restoring diffeomorphism invariance,
 or the geometry must be restricted so that only spacetimes
 with a vanishing Pontryagin density,
 ${^*R}R =0$, are allowed.
 
 This behavior,
 that the consistency of the theory requires
 the condition ${^*R}R =0$, 
 was noted and discussed in Ref.\ \cite{rjsp}.
 Here, however,
 it is used to illustrate how the potential
 inconsistency between dynamics and geometry 
 due to explicit diffeomorphism breaking
 can in some cases be evaded by restricting the possible geometry that can occur.

\subsection{Multiplicative Background Scalars}

In addition to the two specific examples described above, 
theories with a multiplicative background scalar 
can be examined in general as well.

Consider a symmetry-breaking term of the form
${\cal L}_{\rm LV} = \varphi (x) {\cal F}$.
Here, $\varphi (x)$ is a prescribed nondynamical scalar background and ${\cal F}$ is an
arbitrary dynamical scalar function constructed from the metric and its derivatives.
For example ${\cal F}$ could consist of products of contractions of the
curvature tensor.
Variation of the metric in a theory with a term of this form defines a tensor
$ T^{\mu\nu}_{\rm LV}$ as in \rf{geq1},
where integrations by parts are used.
Consistency with Einstein's equations and the contracted Bianchi identity
then requires that $D_\mu T^{\mu\nu}_{\rm LV} = 0$ must hold on shell.

At the same time, the term with ${\cal L}_{\rm LV}$ is a scalar under
observer diffeomorphisms,
obeying $(\de S_{\rm LV})_{\rm observer}  =  0$.
This leads to a condition that can be written as
 \beq
\int d^4x \sqrt{-g}  \,
\xi_\nu \left( D_\mu T^{\mu\nu} -  \,  (D^\nu \varphi) {\cal F}  \right) = 0 ,
 \label{ScalarobsLV}
 \eeq
 which must hold for arbitrary $\xi_\nu$.
 Here, the integrand in \rf{ScalarobsLV}
 does not equal a total derivative.
 This is because the Lie derivative ${\cal L}_\xi \varphi$ 
 only involves factors of $\xi^\mu$ and there are no derivatives acting on $\xi^\mu$.
 This is different from the expression in \rf{Ueq2}, 
 which holds for potentials made out of vectors or tensors,
 where the Lie derivatives give rise to derivatives acting on $\xi_\nu$.
 In the absence of a total derivative term in \rf{ScalarobsLV},
 the expression in parentheses must vanish,
and therefore
\beq
D_\mu T^{\mu\nu} =  (D^\nu \varphi ) {\cal F} .
\label{DTvarphiF}
\eeq
This means that on shell there are only two possibilities for 
this type of theory.
Either $D^\nu \varphi = 0$ and the symmetry is restored
or ${\cal F} = 0$ and the geometry is restricted.

In the example of a spacetime-dependent cosmological constant,
$\varphi (x) = \La (x)$ and ${\cal F} = -1/\ka \ne 0$.
Thus, the only option is that $D^\nu \La(x) = 0$ must hold.
However, in Chern-Simons gravity,
$\varphi (x) = \th (x)$ and ${\cal F} = (1/\sqrt{-g}) {^*R}R$.
In this case,
there is a nontrivial option,
which is that geometrically ${^*R}R = 0$ must hold.

As an additional example, consider a symmetry-breaking term
given as ${\cal L}_{\rm LV} = (-1/2 \ka) \bar u(x) R$,
where $\bar u(x)$ is a background scalar and $R$ is the curvature scalar.
This term has the form of one of the leading-order 
terms in the gravity sector of the SME,
which is discussed in more detail in Section V.
Here, the possibility that $\bar u(x)$ can appear as a fixed
background scalar that explicitly breaks diffeomorphisms is considered.
Matching to the above expressions gives $\varphi = \bar u(x)$ and 
${\cal F} = (-1/2 \ka) R$,
and the result that follows is that 
$D_\mu T^{\mu\nu} =  (-1/2 \ka) R (D^\nu \bar u)$.
Thus, when the theory is on shell,
either $R=0$ or $D^\nu \bar u(x) = 0$ must hold.
Since the gravity sector of the SME is intended for phenomenological tests
in curved spacetimes with matter,
the restriction to spacetimes with $R=0$ is not desirable.
Thus, the restriction that $\bar u$ must equal a constant 
would need to be imposed.  
In that case, however, the factor of $\bar u$ can simply be absorbed by a 
redefinition of the coupling in the Einstein-Hilbert action
and it would have no observable effects.

\subsection{Massive Gravity}

In gravity, it is not possible to form a conventional mass term for the graviton using only the metric 
since scalar quadratic products of $g_{\mu\nu}$ are equal to constants.
To avoid this problem, 
in their original formulation of a massive gravity theory,
Fierz and Pauli (FP) used a perturbative approach in Minkowski spacetime,
creating mass terms out of metric excitations instead of the metric itself
\cite{FP39}.  
However, in massive gravity theories in curved spacetime that 
go beyond the perturbative level,
a background tensor is introduced,
which can be denoted generically as a symmetric two-tensor $\bar f_{\mu\nu}$.
The mass terms for the metric are are then formed out of products of $g_{\mu\nu}$
contracted with $\bar f_{\mu\nu}$.

In general, these types of massive gravity models,
are known to suffer from the presence of a ghost mode,
known as a Boulware-Deser ghost,
as well as difficulty in merging with GR in the massless limit
\cite{vDVZ,BD72}.
Only recently have models been found
that seem promising in being able to avoid these issues.
They are known as de Rham, Gabadadze, Tolley (dRGT)
massive gravity models.
The mass term in dRGT gravity
is generated from a potential ${\cal U}(g_{\mu\nu}, \bar f_{\mu\nu})$ that is 
constructed in a way that eliminates the Boulware-Deser 
ghost mode to all orders in a nonlinear treatment
\cite{dRGT,HR,MGreviews}.
In the original dRGT models,
a Minkowski tensor, $\bar f_{\mu\nu} = \et_{\mu\nu}$,
is used as the background field.
It has also been shown that ghost-free theories can be
obtained using a more general background, $\bar f_{\mu\nu}$,
which is different from the Minkowski tensor
\cite{HR}.

Since $\bar f_{\mu\nu}$ is a fixed background tensor,
particle diffeomorphisms are explicitly broken,
and variations of the mass term with respect to these transformations
give the general off-shell condition,
\beq
 \int d^4x \sqrt{-g}  \,
 \fr {\de  {\cal U}} {\de\bar f_{\mu\nu}} {\cal L}_\xi \bar f_{\mu\nu} \ne 0 .
 \label{Uconflictcondition}
 \eeq
 At the same time,
 the theory is generally covariant,
 and therefore under observer diffeomorphisms,
 a second off-shell condition is
 \beq
 \int d^4x \sqrt{-g}  \, 
 \left[ (D_\mu T^{\mu\nu}_{\rm LV}) \xi_\nu
 +  \fr {\de  {\cal U}} {\de \bar f_{\mu\nu}} {\cal L}_\xi \bar f_{\mu\nu}
 \right] = 0.
 \label{Ueq4}
 \eeq
 
 With $D_\mu T^{\mu\nu}_{\rm LV} = 0$ holding on shell, 
 a potential inconsistency arises between \rf{Uconflictcondition}
 and \rf{Ueq4}.
 However, as shown in the previous section,
 the two terms in the integrand differ off shell by a total derivative term,
\beq
(D_\mu T^{\mu\nu}_{\rm LV}) \xi_\nu 
+ \fr {\de  {\cal U}} {\de \bar f_{\mu\nu}} {\cal L}_\xi \bar f_{\mu\nu}
= D_\mu \left(2 \fr {\de {\cal U}} {\de g^{\al\be}} g^{\mu\al} \xi^\be \right)  .
\label{DTU4}
\eeq
As long as the total derivative does not vanish,
the condition in \rf{Ueq4} can hold on shell.
In this case,
the integral in \rf{Uconflictcondition} vanishes on shell as well,
while off shell the theory remains diffeomorphism violating. 

When exact solutions exist in massive gravity,
it is because the would-be-gauge modes 
are able to impose the on-shell condition 
$D_\mu T^{\mu\nu}_{\rm LV} = 0$ 
as a result of their equations of motion.  
In general,
these modes are able to appear in $ T^{\mu\nu}_{\rm LV}$,
since the contractions of $\bar f_{\mu\nu}$ with $g_{\mu\nu}$ 
can involve all ten components of the metric.
However, if an ansatz form for the metric is used,
as in cosmology when the universe
is assumed to be spatially homogeneous and isotropic,
then there may not be enough degrees of freedom in the metric to satisfy
the consistency requirements.
For example, with $\bar f_{\mu\nu} = \et_{\mu\nu}$
and using a spatially flat metric in a
homogeneous and isotropic universe,
it has been shown that no exact solution for dRGT gravity exists
\cite{AdRDGPT11}.
However, by using other forms for the background and metric,
which introduce one or more additional components to work with,
exact solutions describing a spatially flat homogeneous and isotropic
universe have been obtained
\cite{GLM}.

In dRGT massive gravity with $\bar f_{\mu\nu} = \et_{\mu\nu}$,
additional considerations arise because a Minkowski background
leaves invariant a subset of diffeomorphisms with vectors $\xi^\mu$ equal to constants.
For these vectors, the Lie derivatives ${\cal L}_\xi \et_{\mu\nu} = 0$.
As a result, one of the terms in \rf{DTU4} is removed,
and the metric must obey the condition in \rf{currents} on shell, 
which can lead to restrictions on the allowed geometry.
For example,
in a Robertson-Walker model with a spatially flat homogeneous and isotropic metric
that depends on a scale parameter $a(t)$,
\rf{currents} is only satisfied if $a(t)$ is constant,
which is the same result as in \cite{AdRDGPT11}.

While massive gravity models can satisfy the consistency conditions,
the resulting structure of the theory is very different from GR and
theories with diffeomorphism invariance.  
In particular,
the backgrounds $\bar f_{\mu\nu}$ are fixed nondynamical tensors
that are inserted by hand into the theory,
and they are unable to undergo back reactions.
Covariant energy-momentum conservation is only maintained
by the appearance of the extra modes in the metric,
which act as a buffer between the fixed background 
and the other fields in the theory.  
The natural interplay between geometry and matter that occurs in GR
is disrupted since the background $\bar f_{\mu\nu}$ must remain fixed.

Note that in massive gravity,
it is common to promote the background fields to dynamical fields by
introducing four scalar fields, $\ph^a$, with $a=0,1,2,3$,
known as St\"uckelberg fields.
The background is rewritten as $\bar f_{\mu\nu} = D_\mu \ph^a D_\nu \ph^b f_{ab} (\ph)$,
where $f_{ab}$ is defined so that when $D_\mu \ph^a = \de^a_\mu$,
the fixed background $\bar f_{\mu\nu}$ is reproduced.
By having the fields $\ph^a$ transform as scalars under diffeomorphisms,
this restores the diffeomorphism invariance.
However, the number of degrees of freedom is unaltered,
since adding the four new fields offsets the addition of the four symmetries.
Moreover, by choosing a gauge such that $D_\mu \ph^a = \de^a_\mu$,
the equations of motion reduce to the same set as in the explicit-breaking case.
While the use of St\"uckelberg fields may appear to
restore some of the natural features of GR,
the physical nature of these additional fields remains somewhat contrived.
Similar to the would-be-gauge modes,
these fields do not have have direct interactions with matter,
and they largely act as a kind of camouflage for the fixed background.  

There are, however, alternative approaches to massive gravity 
which do not explicitly break diffeomorphisms and which do not 
introduce extra fields.
These are models that spontaneously break Lorentz and
diffeomorphism invariance,
where the massive modes for the metric arise as Higgs excitations
or through a Higgs mechanism.
In this case,
the background fields are vacuum expectation values,
which arise dynamically,
and diffeomorphism invariance still holds in the action
due to the presence of NG modes.
With spontaneous breaking,
the four extra modes in the metric remain purely gauge,
and it is the dynamics of the observable matter fields
that ensures covariant conservation of energy-momentum.
There are also no potential conflicts between the dynamics and
geometrical identities,
and natural back reactions between the geometry and matter fields occur.
Thus, in these approaches,
many of the natural features in GR still hold.  
For further discussions of gravitational Higgs approaches,
see, for example, Refs.\ \cite{ks,rbak,Higgs}
and the references therein.

\subsection{Spontaneous Diffeomorphism Breaking}

As a final example,
gravitational theories with spontaneous local Lorentz and diffeomorphism breaking
can be examined and compared with theories where the breaking is explicit.   
In theories with spontaneous breaking,
there is typically a potential term ${\cal U}$ in the Lagrangian
that is a function of the metric and an additional tensor field,
which induces nonzero vacuum expectation values.
Examples of models with spontaneous spacetime symmetry breaking
include bumblebee models
\cite{ks,akgrav04,bbmodels},
cardinal models
\cite{cmodels},
and models with an antisymmetric two-tensor
\cite{phon}.
The symmetry breaking occurs when nonzero vacuum values
are formed for the tensor and metric fields,
which minimize the potential ${\cal U}$.
The vacuum values can be denoted generically as $ \bar k_\ch$ and $\bar g_{\mu\nu}$.
For simplicity, and for comparison with explicit-breaking models,
it is assumed that the kinetic terms for $k_\ch$
vanish in the vacuum solution.

In general, the potentials ${\cal U}$ consist of functions of a finite number of
independent scalars $X_i$ that are
formed out of contractions of the background tensor and the metric.
The number of possible values of $i$ depends on the type of tensor.
For example,
with a vector background,
there is only one independent scalar, $X$,
and the potential is a function ${\cal U} (X)$.
However,
if the potential is formed out of the metric and  
a symmetric two-tensor,
the potential ${\cal U}$
can be shown to consist of a function of four independent scalars,
$X_1$, $X_2$, $X_3$, and $X_4$.
These are given as traces of products made out of the metric and 
the background tensor.

Denoting the generic form of the potential as ${\cal U} (X_i)$,
the tensor $T^{\mu\nu}_{\rm LV}$ can be computed
using Euler-Lagrange variations,
and the consistency condition $D_\mu T^{\mu\nu}_{\rm LV} = 0$ 
can be written as
\bea 
\sum_i \left[ \left( \fr {\de^2 {\cal U}(X_i)}{\de X_i^2}\right)(D_\mu X_i) \fr {\de X_i}{\de g_{\mu\nu}}
\quad\quad\quad\quad\quad\quad
\right.
\cr
\left.
+ \fr {\de {\cal U}(X_i)} {\de X_i} 
\left( \fr 1 2 g^{\mu\nu} (D_\mu X_i) + D_\mu \left(\fr {\de X_i}{\de g_{\mu\nu}} \right) \right)
\right]
=0 
\label{Ueq}
\eea
Using this expression,
some of the key differences between the processes of 
explicit and spontaneous diffeomorphism breaking
can be examined.

With explicit breaking,
the tensor $ \bar k_\ch$ is a fixed background
that breaks particle diffeomorphism invariance.
In this case, the four equations in \rf{Ueq} become the four 
equations of motion for the would-be-gauge modes in the metric.
These equations combine with the remaining independent Einstein equations 
to determine the full set of metric components.
The scalars $X_i$ contain the metric excitations,
and in order for massive modes to appear there need to be quadratic 
dependence on these terms in ${\cal U}$.
Thus, 
the first and second variations
of the potential ${\cal U}$ with respect to the scalars $X_i$ 
in general need not vanish for the massive excitations.  

In contrast,
when the symmetry breaking is spontaneous
the vacuum solutions $ \bar k_\ch$ and $\bar g_{\mu\nu}$
minimize the potential, and  
\beq 
\fr {\de {\cal U}(X_i)} {\de X_i} = 0 
\label{Umin}
\eeq
holds for these solutions.
Thus, only the first set of terms remains in \rf{Ueq},
which obeys
\beq 
\sum_i  \left( \fr {\de^2 {\cal U}(X_i)}{\de X_i^2}\right)(D_\mu X_i) \fr {\de X_i}{\de g_{\mu\nu}}
=0 .
\label{1stterm}
\eeq
For the vacuum solution to be an extremum of the potential ${\cal U}$,
the second variations of ${\cal U}$ with respect to $X_i$
should in general be nonzero.
The variations of $X_i$ with respect to the metric need not vanish either.
Thus, a general vacuum solution that satisfies \rf{1stterm} holds when the four scalars 
obey $D_\mu X_i = 0$.
As a result,
the potential ${\cal U}(X_i)$ for the vacuum solution
can only consist of scalar combinations of 
$ \bar k_\ch$ and $\bar g_{\mu\nu}$ 
that are constants and have no explicit spacetime dependence.

In addition to the vacuum solutions,
theories with spontaneous symmetry breaking can also have
massless NG modes and massive Higgs modes.
The NG modes are excitations generated by the broken symmetry that stay in the
degenerate vacuum, obeying \rf{Umin}.
Since the broken symmetries are diffeomorphisms,
the vacuum scalars $X_i$ remain constant for the NG excitations.
Thus, the NG modes satisfy $D_\mu X_i = 0$ as well.  
However, the massive Higgs modes are not generated by diffeomorphisms.
They therefore do not have to obey \rf{Umin}
and the scalars $X_i$ need not remain constant.
In this case,
the solutions to \rf{Ueq} are nontrivial
and depend on the form of ${\cal U}$ and the nature of
the tensor fields $k_\ch$.

\section{Post-Newtonian Limit}

The SME is used in phenomenological investigations of Lorentz
and diffeomorphism violation in gravity and particle physics.
The full SME includes both power-counting renormalizable and
nonreormalizable operators
\cite{akmm,qbakrx}.  
Restrictions to subset models can be defined 
for the gravity sector
\cite{akgrav04,qbak06,akjt,qbakrx}, 
quantum electrodynamics
\cite{smeqed}, 
and both relativistic and
nonrelativistic quantum mechanics
\cite{akcl,rbaknrcl}.

To examine gravitational experiments that test for corrections to GR,
a post-Newtonian limit of the SME has been developed
\cite{qbak06}.
It has been used in analyses of data obtained from
lunar laser ranging
\cite{lrange}, 
atom interferometry
\cite{atomint},
short-range gravitational tests
\cite{srgrav},
analyses of baryon number asymmetry
\cite{GL06},
satellite ranging
\cite{Iorio}, 
gyroscope precession
\cite{gyro},
pulsar timing
\cite{pulsar}, 
and perihelion and solar-spin tests
\cite{qbak06,Iorio}.
These experiments have achieved sensitivities to Lorentz violation
down to levels on the order of parts in $10^{10}$.

\subsection{Spontaneous Breaking}

In the gravity sector of the SME,
it is assumed that the SME coefficients arise
through a process of spontaneous symmetry breaking.
The SME coefficients in this case are vacuum expectation values,
and the theory also must also account for the NG modes and massive Higgs modes
associated with the symmetry breaking.

The development of the post-Newtonian limit of the SME is
described in detail in Refs.\ \cite{qbak06,qbakrx}.
Here, only a brief summary is given.
The starting point is the action in the gravity sector of the SME
in Riemann spacetime.
It consists of three terms written as
\beq
S = S_{\rm EH} + S_{\rm LV} + S^\prime .
\label{Ssme}
\eeq
The first is the Einstein-Hilbert action.
The second contains the interactions involving
the vacuum values and the gravitational fields.
It consists of a series of covariant gravitational operators
of increasing mass dimension,
\beq
S_{\rm LV} = \fr {1}{2\ka} \int \sqrt{-g} \, d^4x 
({\cal L}_{\rm LV}^{(4)} + {\cal L}_{\rm LV}^{(5)} + {\cal L}_{\rm LV}^{(6)} + \cdots) ,
\label{llvsum}
\eeq
where the superscripts denote the mass dimension.

The leading-order terms are of dimension four and have the form
\beq
{\cal L}_{\rm LV}^{(4)}  = -u R 
+\sss R^T_\mn + \ttt C_{\ka\la\mu\nu} ,
\label{llv}
\eeq
where
$R^T_\mn$ is the trace-free Ricci tensor 
and $C_{\ka\la\mu\nu}$ is the Weyl conformal tensor.  
The coefficients $u$, $s^\mn$ and $t^{\ka\la\mu\nu}$
are dynamical fields that acquire vacuum values denoted
as $\bar u$, $\bar s^\mn$ and $\bar t^{\ka\la\mu\nu}$
in a process of spontaneous local Lorentz and diffeomorphism breaking.
The coefficients $s^\mn$ and $t^{\ka\la\mu\nu}$
have symmetries that match those of
the Ricci tensor and the Riemann curvature tensor,
respectively.  
The coefficients $s^\mn$ are traceless 
and the all of the traces of $t^{\ka\la\mu\nu}$ vanish.  

In a process of spontaneous symmetry breaking 
the fields $u$, $s^\mn$, and $t^{\ka\la\mu\nu}$
consist of their background vacuum values 
as well as small fluctuations denoted using tildes,
\bea
u &=& \ub + \utw,
\nonumber\\
\sss &=& \sb^\mn + \stw^\mn,
\nonumber\\
\ttt &=& \tb^{\ka\la\mu\nu} + \ttw^{\ka\la\mu\nu}.
\label{texp}
\eea
The fluctuations include the NG excitations and massive modes.  
The third term in the action, $S^\prime$,
describes the dynamics of these excitations as well
as the dynamics of the ordinary matter fields.
It includes the the kinetic terms for the fluctuations
$\tilde u$, $\tilde s^\mn$, and $\tilde t^{\ka\la\mu\nu}$.

In the post-Newtonian limit,
spacetime is assumed to be asymptotically flat,
and the metric is expanded perturbatively to first order around a Minkowski metric $\et_{\mu\nu}$.
To obtain results involving only gravity corrections,
the excitations in the fields $u$, $s^\mn$, and $t^{\ka\la\mu\nu}$ must be decoupled and 
eliminated in terms of the metric fluctuations.
In general it might seem that this is not possible without giving definite expressions for
the action term $S^\prime$.
However,
as described in Ref.\ \cite{qbak06},
by making a series of assumptions and by exploiting the diffeomorphism invariance,
a general post-Newtonian expansion involving only the vacuum values 
$\bar u$, $\bar s^\mn$, and $\bar t^{\ka\la\mu\nu}$ can be obtained.
Central amongst these assumptions is that the contracted Bianchi identities hold.
These combined with conditions stemming from the diffeomorphism invariance 
of the linearized theory allow the
background fluctuations, $\tilde u$, $\tilde s^\mn$, and $\tilde t^{\ka\la\mu\nu}$,
to be decoupled from the vacuum values and metric excitations.  
The result is a post-Newtonian expansion
involving only the metric, the SME coefficients, $\bar u$, $\bar s^\mn$, and $\bar t^{\ka\la\mu\nu}$,
and the relevant parameters describing a given self-gravitating system.  

Recently, the post-Newtonian limit including gravitational operators of dimension
five and six has been worked out as well
\cite{qbakrx}.
At dimension five, the operators in ${\cal L}_{\rm LV}^{(5)}$
consist of terms having the form of covariant
derivatives acting on the curvature tensor, $D^\la R^{\al\be\ga\de}$.
However, this type of term is CPT odd and would represent pseudovector contributions
to the Newtonian gravitational force rather than conventional vector contributions.
It therefore does not have any effects on nonrelativistic gravity.
The dimension-six operators contribute terms of the form
\bea
{\cal L}_{\rm LV}^{(6)} =
\left( \half (k_1^{(6)})_{\al\be\ga\de\ka\la} 
(D^\ka D^\la R^{\al\be\ga\de} + D^\la D^\ka R^{\al\be\ga\de}) \right.
\nonumber
\\
\left.
+ (k_2^{(6)})_{\al\be\ga\de\ka\la\mu\nu} R^{\ka\la\mu\nu} R^{\al\be\ga\de} \right),
\quad\quad
\label{llv6}
\eea
where the indices on the coefficients $(k_1^{(6)})_{\al\be\ga\de\ka\la} $
and $(k_2^{(6)})_{\al\be\ga\de\ka\la\mu\nu}$ have symmetries that match
the operators that they multiply.  
The vacuum values for the coefficients are denoted as
$(\bar k_1^{(6)})_{\al\be\ga\de\ka\la} $
and $(\bar k_2^{(6)})_{\al\be\ga\de\ka\la\mu\nu}$.
As described in \cite{qbakrx},
a procedure for eliminating the fluctuations in the coefficients about
their vacuum values has been worked out,
and a post-Newtonian limit including contributions from these 
higher-dimensional terms has been obtained as well.

\subsection{Explicit Breaking}

This procedure for finding the post-Newtonian limit of the SME clearly depends on the 
assumption that the local Lorentz and diffeomorphism breaking is spontaneous.
The possibility of explicit diffeomorphism breaking is not considered in \cite{qbak06}
due to the conflicts that can arise between the dynamics and geometrical identities.
However,
in light of the fact that there do exist gravitational theories with explicit breaking
that evade these conflicts,
it is appropriate to examine whether a gravity sector of the SME
with explicit breaking can be defined in a consistent way.  

To modify the gravity sector of the SME for the case of explicit breaking,
the action is assumed to depend on the fixed background values from the start.
The leading-order contributions come from the
dimension-four terms in ${\cal L}_{\rm LV}^{(4)}$,
which in the case of explicit breaking have the form
\beq
{\cal L}_{\rm LV}^{(4)} = -{\bar u} R 
+\bar s^{\mu\nu} R^T_\mn + \bar t^{\ka\la\mu\nu} C_{\ka\la\mu\nu} .
\label{llv}
\eeq
Here, 
$\bar u$, $\bar s^\mn$ and $\bar t^{\ka\la\mu\nu}$
are treated as nondynamical background fields that explicitly break diffeomorphisms.
Since there are no excitations in the background fields,
the action $S^\prime$ reduces to terms for the ordinary matter sector,
which can be ignored in the post-Newtonian limit.
Thus,
the Einstein equations in this case become
$G^{\mu\nu} = T^{\mu\nu}_{\rm LV}$,
where $ T^{\mu\nu}_{\rm LV}$ is defined in \rf{geq1}.

Consistency of the theory requires that $D_\mu T^{\mu\nu}_{\rm LV} = 0$ must hold on shell.
In a theory with explicit breaking,
this is possible as long the extra would-be-gauge modes in the metric
have solutions consistent with this condition.
Writing out this expression in terms of the background fields 
$\bar u$, $\bar s^\mn$, and $\bar t^{\ka\la\mu\nu}$
gives the equation:
\bea
-R (D^\nu \bar u) + 2 g^{\mu\nu} (D_\al R_{\mu\be} ) \bar s^{\al\be}
+ 2 g^{\mu\nu} R_{\mu\be} (D_\al  \bar s^{\al\be})
\cr
+ (D^\nu \bar s^{\al\be}) R_{\al\be} 
+ 4 (D_\al R^\nu_{\pt{\nu}\la\mu\ka}) \bar t ^{\al\la\mu\ka}
\quad\quad\quad
\cr
+ 4 R^\nu_{\pt{\nu}\la\mu\ka} (D_\al  \bar t ^{\al\la\mu\ka})
+ R_{\al\be\ga\de} (D^\nu \bar t ^{\al\be\ga\de}) = 0 .
\quad 
\label{condDT0}
\eea
At the nonlinear level,
the would-be-gauge modes can appear in this equation,
and solutions for these modes can be obtained in principle.
The six independent Einstein equations can then be
used to solve for the remaining metric modes.
Thus, in principle a nonlinear gravity sector of the SME 
could be constructed in the case where the symmetry breaking is explicit.

However,
in practice,
the usefulness of the SME stems from the fact that the post-Newtonian limit provides a
framework in which the leading-order corrections to Newtonian gravity due
to Lorentz violation can be computed.
In this limit,
a linearization of the theory is used,
where the metric is written as $g_{\mu\nu} = \et_{\mu\nu} + h_{\mu\nu}$
and the equations of motion are expanded to lowest order in the excitations $h_{\mu\nu}$.
After carrying out the linearization,
the equations in \rf{condDT0} become
\bea
-R (\partial^\nu \bar u) + 2 \et^{\mu\nu} (\partial_\al R_{\mu\be} ) \bar s^{\al\be}
+ 2 \et^{\mu\nu} R_{\mu\be} (\partial_\al  \bar s^{\al\be})
\cr
+ (\partial^\nu \bar s^{\al\be}) R_{\al\be} 
+ 4 (\partial_\al R^\nu_{\pt{\nu}\la\mu\ka}) \bar t ^{\al\la\mu\ka}
\quad\quad\quad
\cr
+ 4 R^\nu_{\pt{\nu}\la\mu\ka} (\partial_\al  \bar t ^{\al\la\mu\ka})
+ R_{\al\be\ga\de} (\partial^\nu \bar t ^{\al\be\ga\de}) \simeq 0 .
\quad 
\label{condDT0linear}
\eea

With this result,
it can be verified directly that the linearized equations are invariant
under diffeomorphisms,
since the linearized curvature tensor and its contractions are gauge invariant.
Thus,
any would-be-gauge modes that appear in $h_{\mu\nu}$
in the form $(\partial_\mu \Xi_\nu + \partial_\nu \Xi_\mu)$ completely drop out.
This means that the four equations in \rf{condDT0linear} must be solved by
imposing restrictions on the curvature tensor.
These restrictions can be obtained
treating each of the background fields
$\bar u$, $\bar s^\mn$, and $\bar t^{\ka\la\mu\nu}$ independently.
For the background $\bar u$,
the result is already given in Section IV C.
There it was shown that either $\bar u$ must be a constant or $R=0$ must hold.
For the tensor backgrounds, $\bar s^\mn$ and $\bar t^{\ka\la\mu\nu}$,
it is not sufficient that these coefficients are constants.
For consistency,
the gravitational excitations at the linearized level
also need to obey 
\beq
\partial_\al R_{\mu\nu} \simeq 0,
\quad\quad
\partial_\al R_{\ka\la\mu\nu} \simeq 0 .
\label{Riszero}
\eeq
Thus, a spacetime with nonconstant curvature is in general incompatible 
with the presence of the explicit-breaking background fields at the linearized level.
The only consistent solutions either set the background fields to zero
or require that the curvature contributions are constant.
It is therefore unlikely that a useful post-Newtonian limit of
the SME exists when the symmetry breaking is explicit.  

Note these conclusions continue to hold when dimension six terms in 
${\cal L}_{\rm LV}^{(6)}$ are included as well.
With explicit breaking,
the SME coefficients $(\bar k_1^{(6)})_{\al\be\ga\de\ka\la} $
and $(\bar k_2^{(6)})_{\al\be\ga\de\ka\la\mu\nu}$ are
treated as fixed backgrounds,
and consistency requires that $D_\mu T^{\mu\nu}_{\rm LV} = 0$
must hold with these terms included.
In the linearized limit in a post-Newtonian framework,
the terms involving $(\bar k_2^{(6)})_{\al\be\ga\de\ka\la\mu\nu}$
have no contributions to first order in the metric fluctuations $h_{\mu\nu}$.
However, the terms involving $(\bar k_1^{(6)})_{\al\be\ga\de\ka\la} $
do have contributions linear in $h_{\mu\nu}$.
The resulting conditions are again found to be diffeomorphism invariant
in the linearized case.
Repeating the same line of reasoning as above,
the consistency conditions can be shown to hold provided
the background coefficients are constant and
the curvature tensor has vanishing derivatives,
\beq
\partial_\al \partial^\mu \partial^\nu R^{\al\be\ga\de} \simeq 0,
\quad\quad
\partial_\mu \partial^\mu \partial^\nu R^{\al\be\ga\de} \simeq 0 .
\label{Riszero}
\eeq
Thus, the same conclusion holds with explicit breaking stemming 
from dimension-six operators.
A spacetime with nonzero curvature is incompatible with the presence of
explicit-breaking backgrounds in the post-Newtonian limit.

These conclusions for the post-Newtonian limit with explicit breaking 
are in sharp contrast to what happens in the case of spontaneous breaking.
When the symmetry breaking is spontaneous,
there are NG modes that occur in the theory.
These restore the diffeomorphism invariance and there
are no conflicts between the Bianchi identities and covariant energy-momentum conservation.
As a result, there are no restrictions that are imposed on the curvature in the linearized limit. 
Thus,
for the case of spontaneous diffeomorphism breaking,
the post-Newtonian limit of the SME is well defined and can be
used in experimental tests.

\section{Summary \& Conclusions}

The idea that local Lorentz and diffeomorphism invariance might
not hold exactly in physics that goes beyond Einstein's GR
and the Standard Model has been the subject of much theoretical 
and experimental investigation in recent years.
The SME provides the framework used in phenomenological
tests searching for Lorentz and diffeomorphism violation.
The symmetry-breaking terms in the SME
can be probed and measured experimentally,
often with very high sensitivity.
In the gravity sector of the SME, 
a post-Newtonian limit can be derived and used to look for 
departures from Newtonian gravity caused by the spacetime symmetry breaking.

The background fields in the gravity sector of the SME are assumed to arise from a
process of spontaneous symmetry breaking.
However, other modified theories of gravity include fixed
backgrounds that explicitly break local Lorentz symmetry and diffeomorphisms.
Examples of these include Chern-Simons gravity and massive gravity.

This paper has looked at the differences between explicit and
spontaneous diffeomorphism breaking in 
gravitational effective field theories that contain a background field.
In particular, it has been shown that very different 
interpretations hold for the background fields when the
symmetry breaking is explicit versus spontaneous.  
In the case of spontaneous breaking,
the background fields arise as vacuum expectation values.
They are therefore dynamical in nature.
There are also NG modes associated with the symmetry breaking.
However,
when the breaking is explicit,
the background fields are nondynamical,
the diffeomorphism invariance is destroyed from the outset,
and there are no NG modes.

A central feature that distinguishes explicit breaking from spontaneous
breaking of diffeomorphisms is that a potential conflict arises between
the dynamics and geometrical identities for the case of explicit breaking,
while these conflicts are avoided when the breaking is spontaneous.
In Riemann spacetime,
the conflict arises because even when particle diffeomorphisms
are explicitly broken,
general coordinate invariance must still hold. 
As a result,
mathematical conditions resulting from observer diffeomorphism transformations
lead to a potential conflict between the on-shell
Einstein equations and the Bianchi identities.
For theories with covariantly coupled conventional matter,
the consistency condition becomes that $D_\mu T^{\mu\nu}_{\rm LV} = 0$
must hold even when off-shell diffeomorphism invariance is lost.

At the same time, however, when diffeomorphisms are explicitly broken,
four extra independent degrees of freedom appear in the metric.
These are the degrees of freedom that would be gauge in
a theory with diffeomorphism invariance,
such as GR or a theory with spontaneous breaking.
With explicit breaking,
the would-be-gauge components have independent equations of motion
that impose the condition $D_\mu T^{\mu\nu}_{\rm LV} = 0$,
which can allow the potential conflict to be evaded.

Examples of how this occurs or does not occur have been examined.
In certain cases, the conflict is avoided when the integrand in the
variation of the action due to the broken particle diffeomorphisms
equals a total divergence.
However,
in other cases,
the total divergence is absent and the conflict remains.
As a result,
either geometrical restrictions must be imposed, 
or the theory is ruled out.  
For example,
in Chern-Simons gravity,
the consistency condition requires that the spacetime
must have a vanishing Pontryagin density,
otherwise it is inconsistent.
Similarly,
in massive gravity,
there are ansatz solutions for the metric that
cannot be reconciled with the consistency conditions
for particular choices of the background field.  

In the gravity sector of the SME,
since it is assumed that the diffeomorphism breaking is spontaneous,
there are no potential conflicts.
However, since theories with explicit breaking have been found to 
evade the potential conflicts in a variety of cases,
the question arises as to whether the SME coefficients in
the gravity sector can be
treated as being due to explicit breaking.

To address this question,
the gravity sector of the SME was truncated to include only
explicit-breaking fixed backgrounds.
It was found that in principle a consistent model is possible at the nonlinear level.
However, to be useful in phenomenology, 
a consistent post-Newtonian limit should exist,
which can then be used to investigate experimental tests 
of Lorentz and diffeomorphism violation in the presence of gravity.
The result found here is that in the post-Newtonian limit,
consistency only holds if the curvature tensor is a constant.
This effectively rules out using the post-Newtonian limit in gravity experiments
when the symmetry breaking is explicit.  
In contrast, however,
when the symmetry breaking is spontaneous,
the post-Newtonian limit is known to be consistent,
and it has been used in numerous experimental tests with gravity.

Lastly, it is important to keep in mind that
even when a theory with explicit breaking is able
to evade the potential inconsistency,
its resulting structure is fundamentally different from GR,
and many of the compelling features that occur in GR are lost.  
For example,
when diffeomorphism invariance holds,
there is a natural link between the dynamics of the matter fields 
and the geometry of spacetime,
and both the matter and metric fields have natural back reactions with each other.
However,
with explicit breaking, these connections are lost,
since the fixed background is unable to have back reactions
or to exchange energy-momentum density.
To compensate for this,
the extra degrees of freedom in the form of the would-be-gauge modes in the metric must
step in and act as a buffer between the fixed background and the other fields in the theory.
This behavior of the metric is thus very different from GR. 

In contrast,
when diffeomorphisms are spontaneously broken,
many of the natural features of GR are retained
and no absolute objects or prior geometry enter into the theory.  
The background in this case is a vacuum solution of the equations of motion.
The NG modes combined with the vacuum solution maintain diffeomorphism invariance,
and together they have natural back reactions with the other fields in the theory.


\end{document}